\pdfoutput=1
\RequirePackage{ifpdf}
\documentclass{JINST}

\usepackage{subfig}
\title{Reducing beam hardening effects and metal artefacts using Medipix3RX: With applications from biomaterial science}

\author{K. Rajendran$^a$\thanks{Corresponding author.}~, 			
M. F. Walsh$^a$, 
N. J. A. de Ruiter$^a$, 
A. I. Chernoglazov$^b$, 
R. K. Panta$^a$, 
A. P. H. Butler$^{a,b,e,f}$, 
P. H. Butler$^{b,e,f}$, 
S. T. Bell$^f$, 
N. G. Anderson$^a$, 
T. B. F. Woodfield$^{a,b}$, 
S. J. Tredinnick$^b$, 
J. L. Healy$^b$, 
C. J. Bateman$^a$, 
R. Aamir$^a$, 
R. M. N. Doesburg$^{b,e}$, 
P. F. Renaud$^b$, 
S. P. Gieseg$^b$, 
D. J. Smithies$^f$, 
J. L. Mohr$^a$, 
V. B. H. Mandalika$^b$, 
A. M. T. Opie$^b$, 
N. J. Cook$^d$,
J. P. Ronaldson$^a$, 
S. J. Nik$^b$,
A. Atharifard$^a$, 
M. Clyne$^g$,
P. J. Bones $^b$, 
C. Bartneck$^b$, 
R. Grasset$^b$, 
N. Schleich$^c$ 
and M. Billinghurst$^b$ \\
\llap{$^a$}University of Otago,\\
2 Riccarton Ave, Christchurch 8140, New Zealand\\
\llap{$^b$}University of Canterbury,\\
Private Bag 4800, Christchurch 8140, New Zealand\\  
\llap{$^c$}University of Otago,\\
23A Mein Street, Newtown, Wellington 6021, New Zealand\\
\llap{$^d$}Canterbury District Health Board\\
Riccarton Avenue, PO Box 4710, Christchurch 8140, New Zealand\\
\llap{$^e$}The European Organization for Nuclear Research (CERN)\\
Geneva, Switzerland\\
\llap{$^f$}MARS Bioimaging Ltd\\
29a Clyde Rd, Christchurch, New Zealand\\
\llap{$^g$}ILR\\
Christchurch, New Zealand\\
E-mail: \email{kishore.rajendran@canterbury.ac.nz}}
\abstract{This paper discusses methods for reducing beam hardening effects using spectral data for biomaterial applications. A small-animal spectral scanner operating in the diagnostic energy range was used. We investigate the use of photon-processing features of the Medipix3RX ASIC in reducing beam hardening and associated artefacts. A fully operational charge summing mode was used during the imaging routine. We present spectral data collected for metal alloy samples, its analysis using algebraic 3D reconstruction software and volume visualisation using a custom volume rendering software. Narrow high energy acquisition using the photon-processing detector revealed substantial reduction in beam hardening effects and metal artefacts.}  

\begin{document}

\renewcommand{\thefootnote}{\alph{footnote}}

\section{Introduction}

Beam hardening artefacts in x-ray computed tomography (CT) of metal hardware frequently limit the diagnostic quality. This applies to large metal objects such as orthopaedic implants as well as smaller samples like tissue-engineered scaffolds. Beam hardening effects are visible in the form of streak artefacts and cupping effects. These artefacts affect the metal and surrounding non-metal regions. During beam hardening, the mean energy of the x-ray beam increases, and dense metal samples can cause severe beam hardening due to reasonably high atomic number as compared to soft tissues. Photon starvation due to the presence of metal hardware further deteriorates the image quality. One common approach is to employ a source filter and preharden the beam, thus removing low energy photons that largely contribute to beam hardening. However, this leads to poor soft tissue contrast. Mathematical correction techniques for beam hardening and metal artefacts have been reported, some of them used in clinical routines. 
These techniques 
include sinogram interpolation methods \cite{NMAR, FSMAR}, dual energy extrapolation \cite{monoenergetic}, energy models and polynomial corrections (for $\mu$CT) \cite{bimodal, polynomial}. Some of these post-processing techniques are computationally intense, requiring metal data segmentation and/or several forward and backward projections. Dual energy corrections are usually done at the cost of increased exposure \cite{polynomial}.

\begin{figure}[h!]
  \begin{center}  
    \subfloat[15 to 80 keV]{\label{15to80comp}\includegraphics[width=0.35\linewidth]{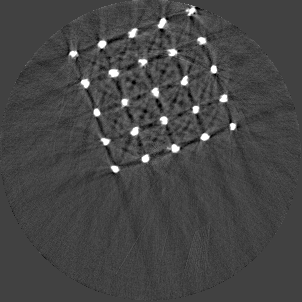}}    
    \subfloat[55 to 80 keV]{\label{55to80comp}\includegraphics[width=0.35\linewidth]{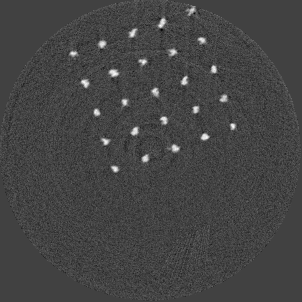}}      
  \end{center}  
  \caption{Metal scaffold sample imaged using Medipix3RX. High energy range shows reduced artefacts compared to wide energy acquisition.}
  \label{compare}
\end{figure}

Photon-counting detectors have been successfully employed in preclinical applications. Using spectral imaging, a novel approach towards minimising beam hardening effects is proposed.  To the best of our knowledge, beam hardening and metal artefact reduction using spectral imaging has not been reported. Unlike numerical techniques, the work described in this paper aims at minimising metal artefacts in the acquisition stage, by capturing high energy quanta that exhibit less beam hardening effects. The Medipix ASIC allows simultaneous data acquisition from discrete user-defined energy ranges. The ASIC was designed to count photon events and categorise them based on energy thresholds determined by the user. This feature enables the capture of spectral signatures for multiple materials which can be used for material discrimination. The number of counts for discrete energy bands can be obtained by subtracting data from two counters. This is essentially done as a pre-processing step prior to flat-field 
normalization 
and reconstruction. The raw data from a counter has an energy range between [$T_{C}$, kVp], where $T_{C}$ is the corresponding user-defined threshold and kVp is the x-ray tube potential used. Since the count information is acquired simultaneously in a single exposure, the noise in a particular energy range is local Poisson noise due to quantum fluctuations. Reduction in streak artefacts using spectral imaging of a scaffold sample is shown in figure \ref{compare}. Wide energy acquisition shows severe streaks while narrow high energy range exhibits reduced artefacts. Also, spatial improvements corresponding to the metal region can be noticed in the narrow energy band while the wide energy reconstruction shows a blooming effect.

The earlier Medipix detectors had technical challenges relating to charge sharing effects which is prominent when pixel pitch less than 300 $\mu$m is used. Charge sharing is when the total charge from a single photon event is split across several pixels and individual pixels counted them as seperate events thereby affecting spatial and spectral resolution. To overcome this problem, the new Medipix3RX enables a fully operational charge summing mode (CSM) \cite{3RX}. During charge sharing, a single event across multiple pixels are identified through inter-pixel communication. The charge from 2 x 2 pixel region is summed and the pixel receiving the highest charge from the event is alloted the summed total charge. This can be termed as `photon-processing'. A chip without CSM would count each of these pixels as a separate photon, each with a reduced energy. In addition to CSM, there is spectroscopic mode, which allows inter-pixel communication to multiplex the 2 counters per pixel 
over a 2 x 2 pixel 
area (110 x 110 $\mu$m$^2$), giving 8 possible counters. When spectroscopic mode and CSM are enabled, the 110 x 110 $\mu$m$^2$ area pixels are checked with their neighbours, summing the charge over an area of 220 x 220 $\mu$m$^2$. This leads to a spatial arrangement of 128 x 128 pixel clusters at 110 $\mu$m pitch with each pixel cluster providing 4 CSM counters (and 4 arbitrated but non-CSM counters). 
High-Z sensor layers like cadmium telluride (CdTe) are preferred for clinical applications due to their ability to operate in wider energy ranges \cite{CdTe}. A silicon (Si) sensor layer is suitable for soft-tissue imaging in small animals, but has poor detection efficiency over the full diagnostic x-ray range. Due to low absorption efficiencies \cite{Medipix3.1}, Si becomes virtually transparent to hard x-rays, which are required for CT of large objects, for metal artefact reduction and for clinical applications involving gold or gadolinum K-edge imaging. The quantum detection efficiency of CdTe is suitable for operation in human diagnostic x-ray range (20 to 140 keV). This article discusses the initial experiments using CdTe sensor in Medipix3RX ASIC for studying beam hardening effects and metal artefact reduction in the x-ray energy range of 15 to 80 keV. Metal samples of titanium (Ti) and magnesium (Mg) alloys \cite{Mg_porous} that are used in tissue engineering research were used in this study. Porous 
scaffolds of these 
metals are usually implanted in bone to study bone ingrowth \cite{boneingrowth1, boneingrowth2}.     
\section{Materials and Methods}
\begin{figure}[h!]
  \begin{center}  
    \subfloat[Ti scaffold]{\label{Ti_pic}\includegraphics[width=0.21\linewidth]{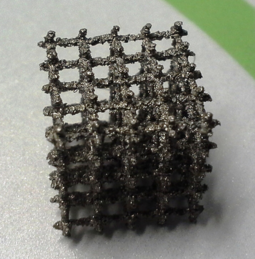}}    
    \subfloat[Mg scaffold]{\label{Mg_pic}\includegraphics[width=0.223\linewidth]{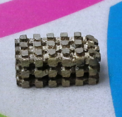}}    
    \subfloat[Ti mesh]{\label{Ti_mesh_pic}\includegraphics[width=0.148\linewidth]{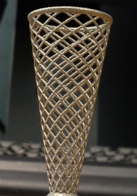}}    
  \end{center}
  \caption{Snapshots of the metal samples (see table 1 for scale information).}
  \label{scaffold_pic}
\end{figure}

For the beam hardening study, we used Medipix All Resolution System (MARS) \cite{firstCT} containing a Medipix3RX ASIC with a 2 mm CdTe sensor bump bonded at 110 $\mu$m in a single chip layout. All the acquisitions in this paper were carried out in CSM. The detector assembly is a module of the MARS camera which also contains a readout board, peltier cooling system and an integrated bias-voltage board. A negative bias voltage of -440 V was applied across the sensor during the acquisitions. The MARS scanner system comprises of MARS camera, a rotating internal gantry and an 80 kVp Source-Ray SB-80-1K x-ray tube (Source-Ray Inc, Ronkonkoma, NY) with a tungsten anode having 1.8 mm aluminium (Al) equivalent intrinsic filtration. The focal spot size is approximately 33 $\mu$m \cite{sourceray}. Mechanical motor control (gantry rotation, source to detector translation, camera translation and sample translation), detector energy response 
calibration and threshold equalization were performed using the custom built MARS scanner software. The samples used in this study are shown in figure \ref{scaffold_pic} and its description is provided in table \ref{sampledesc}. 
\begin{table}[ht]
\caption{Sample description}
\centering
\begin{tabular}{c c p{7cm}}
\hline\hline 
Sample & Material & \centerline{Description}\\ \hline
Metal phantom & Ti alloy & Solid cylinder of 8 mm diameter press fitted onto a Perspex cylinder of 25 mm diameter to study cupping effect. \\
Porous scaffold & Ti alloy & Porous 3D lattice structure fabricated via electron beam melting with $\approx$700 $\mu$m thick struts. Used in tissue engineering research. \\
Porous scaffold & Mg alloy & Porous 3D lattice structure fabricated via an indirect additive manufacturing process in molten Mg with $\approx$500 $\mu$m thick struts. Used in tissue engineering research \cite{Mg_porous}. \\ 
Porous mesh (stent-like pattern) & Ti alloy & Porous 3D structure fabricated via selective laser sintering with variable strut thickness between 620 $\mu$m and 670 $\mu$m. Sample length measures 45 mm (includes a 10 mm base). \\ [1ex] 
\hline
\end{tabular}
\label{sampledesc}
\end{table}

\subsection{Spectral scan parameters}
The scanner gantry was set to continuous-motion rotation to acquire projections at multiple camera positions. The CSM thresholds were set to 15, 35, 55 and 62 keV. All the samples were mounted in air during the spectral scans. The MARS camera was translated vertically to cover the entire diameter of the sample. The geometric parameters (source to detector distance (SDD) and source to object distance (SOD)) and x-ray tube settings are provided in table \ref{settings}.

\begin{table}[ht]
\caption{Scan parameters}
\centering
\begin{tabular}{c c c c c c}
\hline\hline 
    Sample & SOD (mm) & SDD (mm) & Voltage (kVp) & Current ($\mu$A) & Exposure time (ms) \\ [0.5ex]
    \hline
    Ti phantom & 131.8 & 163.8 & 80 & 80 & 50\\ 
    Ti scaffold & 131.8 & 170.8 & 80 & 90 & 40 \\ 
    Mg scaffold  & 131.8 & 170.8 & 80 & 90 & 40  \\ 
    Ti mesh & 110 & 144 & 80 & 80 & 40\\ [1ex]
\hline
\end{tabular}
\label{settings}
\end{table}

\subsection{Post processing chain}
The raw data acquired from the scanner were flat-field normalized using open beam projections (500 open beam projections per camera position per counter acquired prior to scan). Dark-field images (50 dark-fields) were acquired prior to scan for dark-field (bad pixel) correction. A projection space statistical ring filter loosely based on \cite{ring} was applied prior to reconstruction. The projections were reconstructed using MARS-ART (Algebraic Reconstruction Technique) algorithm \cite{tangART}. Number of projections per gantry rotation was set to 720. Volumetric rendering was performed using MARS - Exposure Render, which is a modified version of the open-source Exposure Render \cite{erender} software that implements a direct volume rendering (DVR) algorithm. Modifications to Exposure Render include the addition of tricubic B-spline interpolation between data voxels, the ability to simultaneously visualise up to 8 volumetric datasets, and numerous user interface changes. 
\section{Results and Discussion}
\begin{figure}[h!]
  \begin{center}  
    \subfloat[15 to 80 keV]{\label{15to80P}\includegraphics[width=0.22\linewidth]{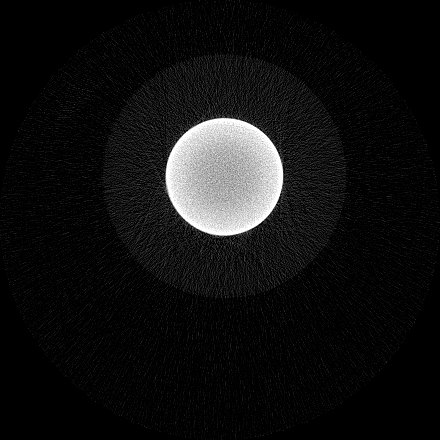}}    
    \subfloat[35 to 80 keV]{\label{35to80P}\includegraphics[width=0.22\linewidth]{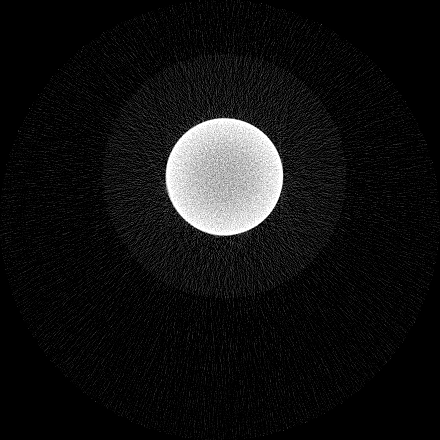}}    
    \subfloat[55 to 80 keV]{\label{55to80P}\includegraphics[width=0.22\linewidth]{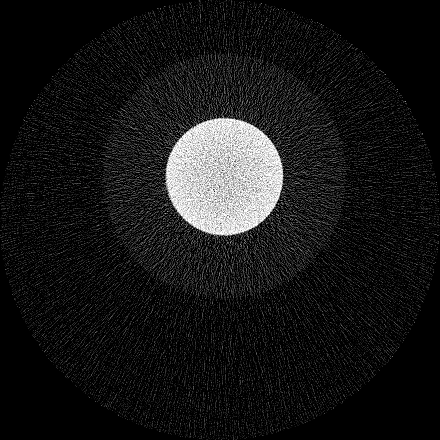}}
    \subfloat[62 to 80 keV]{\label{62to80P}\includegraphics[width=0.22\linewidth]{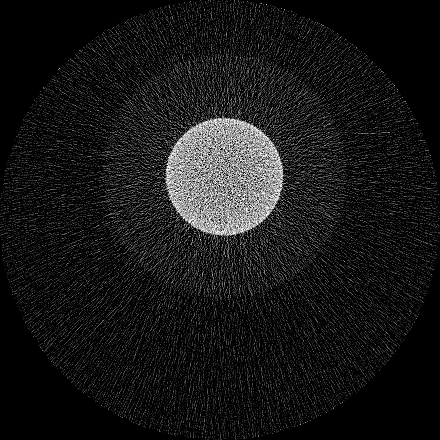}}  
  \end{center}
  \caption{Spectral reconstruction of Ti cylindrical phantom. High energy ranges shows reduced cupping effect.}
  \label{phantom}
\end{figure}

\begin{figure}[h!]
  \begin{center}
    \subfloat[15 to 80 keV]{\label{15to80L}\includegraphics[width=0.52\linewidth]{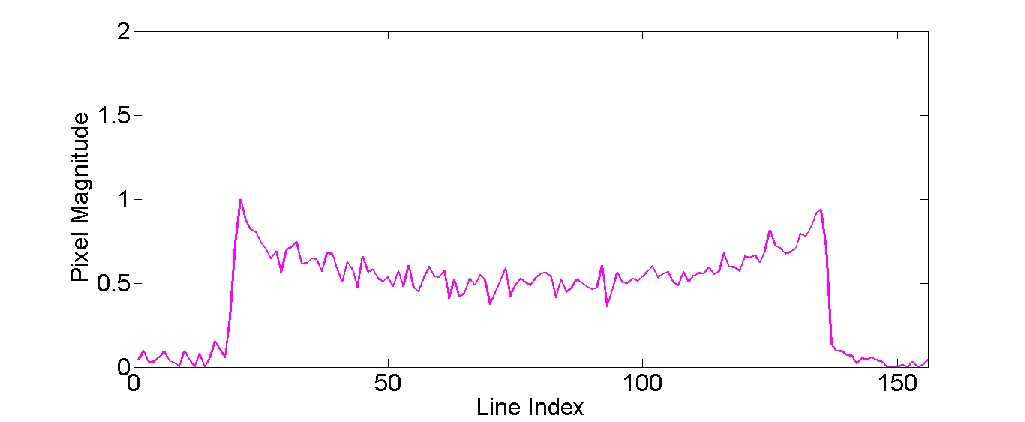}}    
    \subfloat[35 to 80 keV]{\label{35to80L}\includegraphics[width=0.52\linewidth]{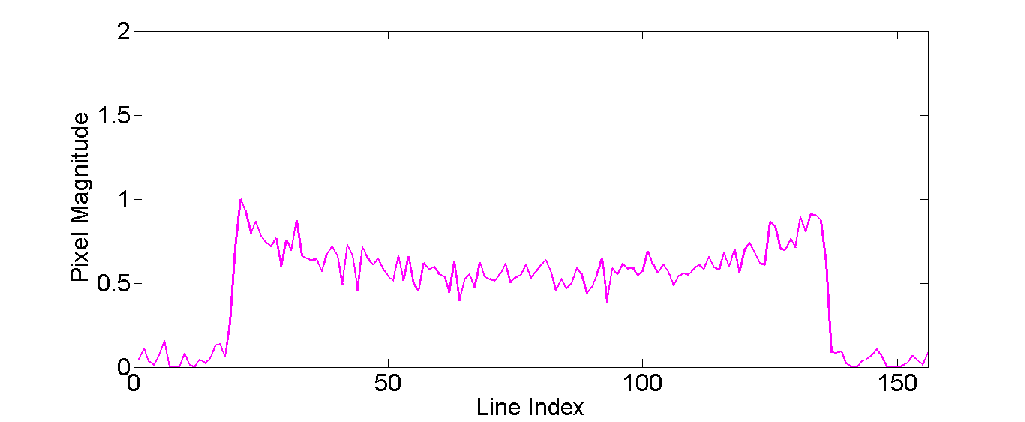}}
    
    \subfloat[55 to 80 keV]{\label{55to80L}\includegraphics[width=0.52\linewidth]{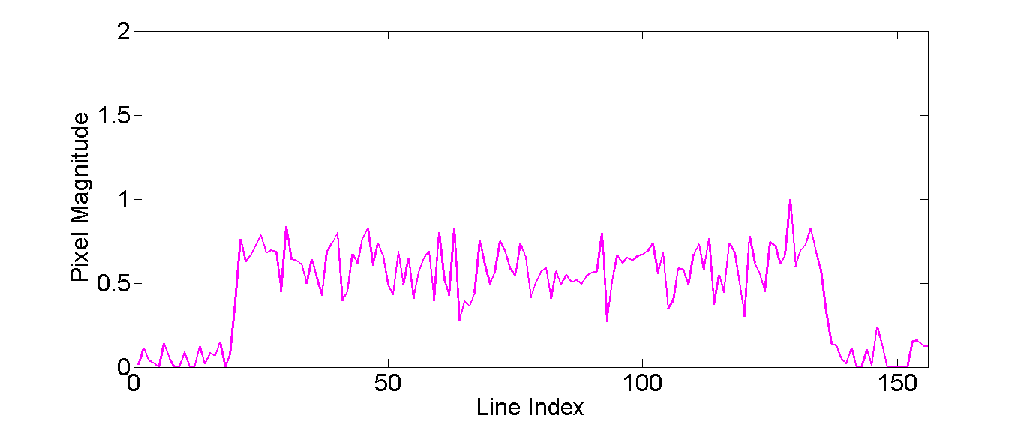}}
    \subfloat[62 to 80 keV]{\label{62to80L}\includegraphics[width=0.52\linewidth]{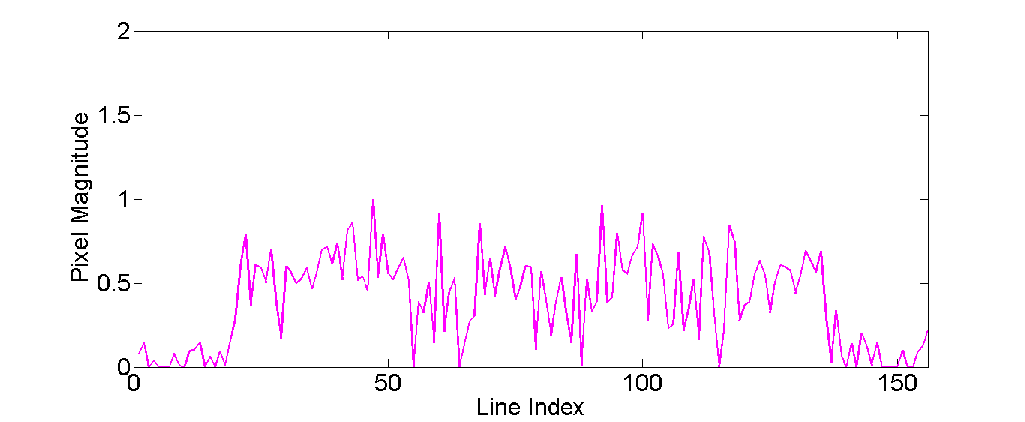}}  
  \end{center}  
  \caption{Normalized line profiles for Ti cylindrical phantom. Cupping effect can be seen in (a) and (b) but much reduced in (c) and (d).}
  \label{line}
\end{figure}

Figure \ref{phantom} shows a single slice spectral reconstruction of the Ti phantom. The cupping effect is prominent in the low energy range and decreases in the high energy acquisitions. The thresholds were determined to provide a trade-off between reduced photon noise and cupping effect. The spectral images for the energy ranges 55 to 80 keV and 62 to 80 keV  exhibit reduced cupping effect while the 15 to 80 keV reconstruction has low quantum noise and shows good contrast in non-metal regions. In figure \ref{line}, a horizontal line profile passing through the origin of the metal cylinder shows cupping effect in the different energy ranges. Without the use of any hardware filters, a significant reduction in the cupping effect is noticeable in figure \ref{55to80L}. The reconstruction corresponding to the energy range from 62 to 80 keV suffers from severe photon limitation giving rise to statistical noise. Any significant increase in tube current and/or exposure time for this scan resulted in detector 
saturation in non-metal regions.  

\begin{figure}[h!]
  \begin{center}
    \subfloat[15 to 80 keV]{\label{roi1}\includegraphics[width=0.25\linewidth]{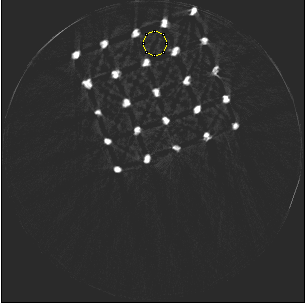}}    
    \subfloat[35 to 80 keV]{\label{roi2}\includegraphics[width=0.25\linewidth]{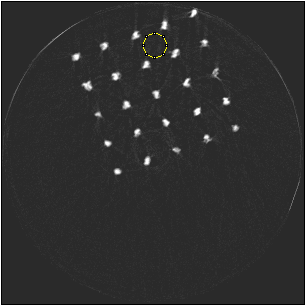}}
    \subfloat[55 to 80 keV]{\label{roi3}\includegraphics[width=0.25\linewidth]{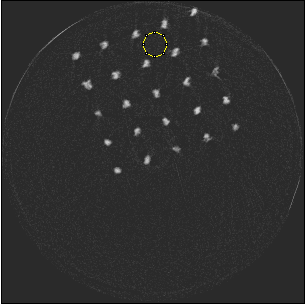}}
    \subfloat[62 to 80 keV]{\label{roi4}\includegraphics[width=0.248\linewidth]{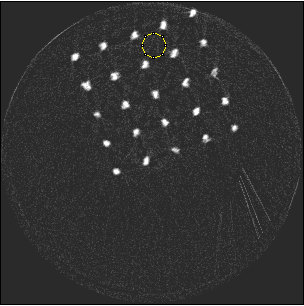}}  
  \end{center}  
  \caption{Single slice spectral reconstruction of Ti scaffold sample. $\mu_{ROI}$ is 0.246, 0.030, 0.008 and 0.103 for the circular ROI in (a), (b), (c) and (d) respectively.}
  \label{Tiscaffold}
\end{figure}

Figure \ref{Tiscaffold} illustrates a single slice spectral reconstruction of the Ti scaffold. Varying levels of streak artefacts can be seen across the spectral reconstructions. The spectral reconstructions for the energy ranges 35 to 80 keV, 55 to 80 keV, and 62 to 80 keV shown in figure \ref{Tiscaffold}, exhibit reduced streak artefacts. A region-of-interest (ROI) analysis was performed in the immediate vicinity of the metal region where the streaks are more pronounced. Average attenuation coefficent of air close to zero conveys less regional noise/artefacts. The regional average attenuation coefficient ($\mu_{ROI}$) of the non-metal (air) region in 55 to 80 keV reconstruction (figure \ref{roi3}) shows reduced artefacts. Even though minor streaks and statistical noise appear in figure \ref{roi4} due to photon limitation, the artefacts are less pronounced in comparison to the wide energy acquisition in figure \ref{roi1}. 

\begin{figure}[h!]
  \begin{center}
    \subfloat[30 to 50 keV using Si]{\label{Si}\includegraphics[width=0.25\linewidth]{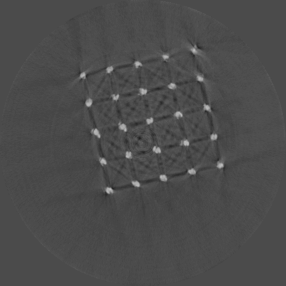}}    
    \subfloat[35 to 55 keV using CdTe]{\label{CdTe}\includegraphics[width=0.25\linewidth]{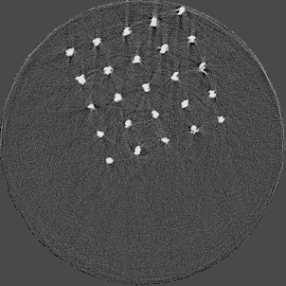}}  
  \end{center}  
  \caption{Ti scaffold reconstruction using Si Medipix3.1 and CdTe Medipix3RX.} 
  \label{Ti_Si_CdTe}
\end{figure}

Using the Ti scaffold sample, a post reconstruction analysis between Si detector (Medipix3.1) operating in Single Pixel Mode (SPM) and CdTe detector (Medipix3RX) operating in CSM was carried out. Figure \ref{Si} shows a reconstruction for energy range 30 to 50 keV obtained using Si detector with a detector element size 55 $\mu$m in SPM. Despite good spatial resolution, artefacts are still prominent. The reconstruction using CdTe detector with a detector element size of 110 $\mu$m in CSM (figure \ref{CdTe}) shows reduced artefacts comparatively. To obtain the narrow energy range of 35 to 55 keV, the raw counts at 35 to 80 keV and 55 to 80 keV were subtracted.

\begin{figure}[h!]
  \begin{center}
    \subfloat[15 to 80 keV]{\label{15to80M}\includegraphics[width=0.25\linewidth]{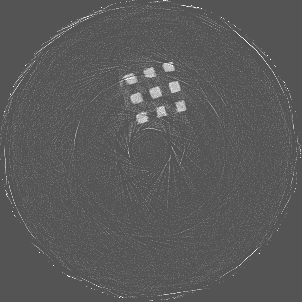}}    
    \subfloat[35 to 80 keV]{\label{35to80M}\includegraphics[width=0.25\linewidth]{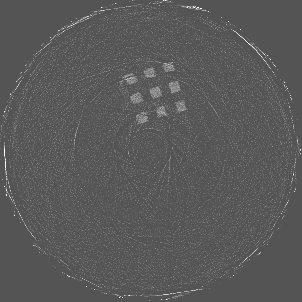}}    
    \subfloat[55 to 80 keV]{\label{55to80M}\includegraphics[width=0.25\linewidth]{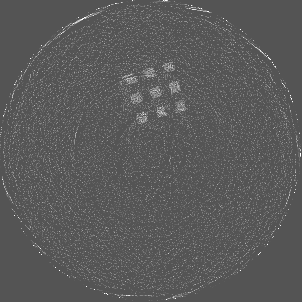}}
    \subfloat[62 to 80 keV]{\label{62to80M}\includegraphics[width=0.25\linewidth]{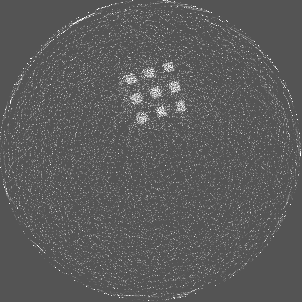}}
  \end{center}  
  \caption{Spectral reconstruction of Mg scaffold. Low energy ranges provide good spatial resolution while high energy ranges are limited by photon noise.}
  \label{fig6:main}
\end{figure}

\begin{figure}[h!]
  \begin{center}
    \subfloat[15 to 80 keV]{\label{15to80v}\includegraphics[width=0.25\linewidth]{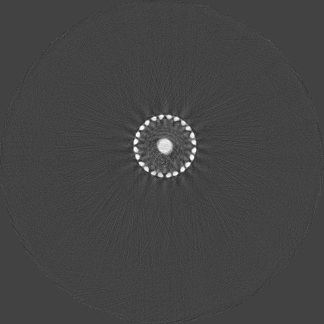}}    
    \subfloat[35 to 80 keV]{\label{35to80v}\includegraphics[width=0.25\linewidth]{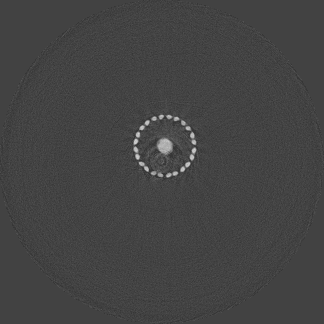}}    
    \subfloat[55 to 80 keV]{\label{55to80v}\includegraphics[width=0.25\linewidth]{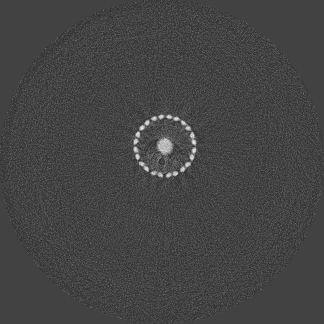}}
    \subfloat[62 to 80 keV]{\label{62to80v}\includegraphics[width=0.25\linewidth]{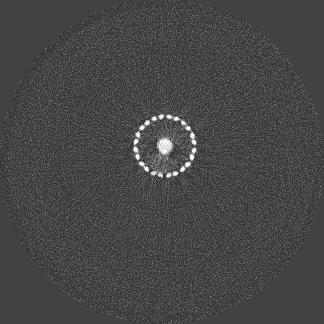}}
  \end{center}  
  \caption{Spectral reconstruction of Ti mesh sample. Minor streaks are visible in the low energy range.}
  \label{meshrec}
\end{figure}

Figure \ref{fig6:main} shows a single slice spectral reconstruction of the Mg scaffold. Due to low atomic number of Mg (Z = 12) compared to Ti (Z = 22), the results did not exhibit any significant beam hardening effects. Low energy reconstruction shows good spatial information while high energy ranges are limited by photon noise. In scans involving smaller samples made from low-Z materials like Al or Mg, acquiring low energy quanta in CSM provide high spatial information with minimum or no beam hardening effects. Figure \ref{meshrec} illustrates a single slice spectral reconstruction of the Ti mesh. Similar to the Ti scaffold, streaks are less pronounced in the mid and high energy ranges.

\section{Summary and Conclusion}
\begin{figure}[h!]
  \begin{center}
    \subfloat[Porous Ti scaffold]{\label{Tiscaff}\includegraphics[width=0.32\linewidth]{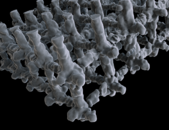}}    
    \subfloat[Mg scaffold]{\label{Mgscaff}\includegraphics[width=0.328\linewidth]{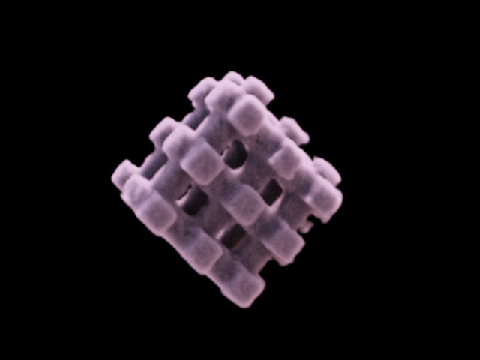}}    
    \subfloat[Ti mesh]{\label{mesh}\includegraphics[width=0.308\linewidth]{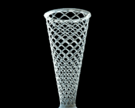}}
  \end{center}  
  \caption{High resolution MARS-Exposure Render visualisation of the metal samples}
  \label{exposnap}
\end{figure}

 Beam hardening and metal artefacts pose challenges during CT imaging in the presence of metal hardware \cite{CTartifact}. This paper presents data that demonstrates the use of spectral imaging in reducing beam hardening effects and metal artefacts. A high-Z sensor layer like CdTe is necessary to provide improved spectral resolution at higher x-ray energies needed for typical implant visualisation. Multi-energy acquisition of metal samples has the added advantage of capturing spectral information which exhibits reduced artefacts and reasonable non-metal (tissue) information. Further, the results were obtained without any hardware filters (except for the intrinsic filter-equivalent in the x-ray tube) and without any numerical corrections. A global reduction in noise due to charge sharing effects was seen due to the availability of CSM. 3D visualisation of the samples (figure \ref{exposnap}) revealed finer spatial structures. 
 
Further improvements in metal artefact reduction should be achieved by (i) increasing the tube potential from 80 kVp to 120 kVp and (ii) numerical correction methods by using a reference energy range where artefacts are minimal. ART plays an important role while reconstructing datasets with low photon counts. Pixels that receive no photons are treated as dead pixels and the corresponding equations are ignored during the algebraic reconstruction, a technique that is not possible in conventional filtered back projection reconstruction. This helps in avoiding secondary artefacts due to incomplete data. Sinograms are not used since MARS-ART directly operates on individual projection frames. Different metal samples and scaffold structures are currently being studied using the MARS spectral imaging modality. Tissue ingrowth quantification using imaging techniques will help identify the biocompatibility of scaffold materials. The quality of information derived from conventional CT imaging are limited by beam 
hardening effects. Spectral imaging has helped identify energy ranges that are less prone to beam hardening effects and provide improved visualisation in the absence of artefacts. In conclusion, high-Z detectors such as CdTe operating in CSM outperform Si detectors for beam hardening and artefact reduction using high energy x-ray range.
The raw data (dicom files), pre-processed projection images and the reconstructions can be obtained from \href{http://hdl.handle.net/10092/8627}{http://hdl.handle.net/10092/8627} for readers to test the data using their familiar routines.
  
\acknowledgments

This project was funded by Ministry of Business, Innovation and Employment (MBIE), New Zealand under contract number UOCX0805. The authors would like to thank all members of MARS-CT project, the Medipix2 collaboration, and the Medipix3 collaboration. In particular we acknowledge the CERN based designers Michael Campbell, Lukas Tlustos, Xavier Llopart, Rafael Ballabriga and Winnie Wong, and the Freiburg material scientists Michael Fiederle, Alex Fauler, Simon Procz, Elias Hamann and Martin Pichotka. We also thank Graeme Kershaw, University of Canterbury for preparing the hardware phantom, and Anton Angelo, University of Canterbury for co-ordinating access to the data repository.

\bibliographystyle{jhep}
\bibliography{paper}

\providecommand{\href}[2]{#2}\begingroup\raggedright\begin{thebibliography}{10}

\bibitem{NMAR}
E.~Meyer, R.~Raupach, M.~Lell, B.~Schmidt, and M.~Kachelrie{\ss}, {\it
  Normalized metal artifact reduction ({NMAR}) in computed tomography},  {\em
  Medical Physics} {\bf 37} (2010), no.~10 5482--5493.

\bibitem{FSMAR}
E.~Meyer, R.~Raupach, M.~Lell, B.~Schmidt, and M.~Kachelrie{\ss}, {\it
  Frequency split metal artifact reduction ({FSMAR}) in computed tomography},
  {\em Medical Physics} {\bf 39} (2012), no.~4 1904--1916.

\bibitem{monoenergetic}
F.~Bamberg, A.~Dierks, K.~Nikolaou, M.~Reiser, C.~Becker, and T.~Johnson, {\it
  Metal artifact reduction by dual energy computed tomography using
  monoenergetic extrapolation},  {\em European Radiology} {\bf 21} (2011),
  no.~7 1424--1429.

\bibitem{bimodal}
E.~Van~de Casteele, D.~Van~Dyck, J.~Sijbers, and E.~Raman, {\it {{A}n
  energy-based beam hardening model in tomography}},  {\em Phys Med Biol} {\bf
  47} (Dec, 2002) 4181--4190.

\bibitem{polynomial}
E.~Van~de Casteele, D.~Van~Dyck, J.~Sijbers, and E.~Raman, {\it A model-based
  correction method for beam hardening artefacts in x-ray microtomography},
  {\em X-Ray Science and Technology} {\bf 12} (2004), no.~1 43--57.

\bibitem{3RX}
R.~Ballabriga, J.~Alozy, G.~Blaj, M.~Campbell, M.~Fiederle, E.~Frojdh, E.~H.~M.
  Heijne, X.~Llopart, M.~Pichotka, S.~Procz, L.~Tlustos, and W.~Wong, {\it The
  {M}edipix3{RX}: a high resolution, zero dead-time pixel detector readout chip
  allowing spectroscopic imaging},  {\em Journal of Instrumentation} {\bf 8}
  (2013), no.~02 C02016.

\bibitem{CdTe}
T.~Takahashi and S.~Watanabe, {\it Recent progress in {C}d{T}e and {C}d{Z}n{T}e
  detectors},  {\em Nuclear Science, IEEE Transactions on} {\bf 48} (2001),
  no.~4 950--959.

\bibitem{Medipix3.1}
M.~F. Walsh, S.~J. Nik, S.~Procz, M.~Pichotka, S.~T. Bell, C.~J. Bateman,
  R.~M.~N. Doesburg, N.~D. Ruiter, A.~I. Chernoglazov, R.~K. Panta, A.~P.~H.
  Butler, and P.~H. Butler, {\it Spectral ct data acquisition with medipix3.1},
   {\em Journal of Instrumentation} {\bf 8} (2013), no.~10 P10012.

\bibitem{Mg_porous}
T.~L. Nguyen, M.~P. Staiger, G.~J. Dias, and T.~B.~F. Woodfield, {\it A novel
  manufacturing route for fabrication of topologically-ordered porous magnesium
  scaffolds},  {\em Advanced Engineering Materials} {\bf 13} (2011), no.~9
  872--881.

\bibitem{boneingrowth1}
P.~Habibovic, T.~Woodfield, K.~Groot, and C.~Blitterswijk, {\it Predictive
  value of in vitro and in vivo assays in bone and cartilage repair - what do
  they really tell us about the clinical performance?},  in {\em Tissue
  Engineering} (J.~Fisher, ed.), vol.~585 of {\em Advances in Experimental
  Medicine and Biology}, pp.~327--360.
\newblock Springer US, 2007.

\bibitem{boneingrowth2}
J.~E. Biemond, G.~Hannink, N.~Verdonschot, and P.~Buma, {\it {{B}one ingrowth
  potential of electron beam and selective laser melting produced
  trabecular-like implant surfaces with and without a biomimetic coating}},
  {\em J Mater Sci Mater Med} {\bf 24} (Mar, 2013) 745--753.

\bibitem{firstCT}
M.~F. Walsh, A.~M.~T. Opie, J.~P. Ronaldson, R.~M.~N. Doesburg, S.~J. Nik,
  J.~L. Mohr, R.~Ballabriga, A.~P.~H. Butler, and P.~H. Butler, {\it First {CT}
  using {M}edipix3 and the {MARS-CT}-3 spectral scanner},  {\em Journal of
  Instrumentation} {\bf 6} (2011), no.~01 C01095.

\bibitem{sourceray}
{Source-Ray Inc.}, {\em Model {SB-80-1K} {(Doc.} M-{SB801K-DI}, Rev 1)
  {Installation/Operation} Manual}.
\newblock 2002.

\bibitem{ring}
J.~Sijbers and A.~Postnov, {\it {{R}eduction of ring artefacts in high
  resolution micro-{C}{T} reconstructions}},  {\em Phys Med Biol} {\bf 49}
  (Jul, 2004) N247--253.

\bibitem{tangART}
N.~D. Tang, N.~de~Ruiter, J.~L. Mohr, A.~P.~H. Butler, P.~H. Butler, and
  R.~Aamir, {\it Using algebraic reconstruction in computed tomography},  in
  {\em Proceedings of the 27th Conference on Image and Vision Computing New
  Zealand}, pp.~216 -- 221, 2012.

\bibitem{erender}
T.~Kroes, F.~H. Post, and C.~P. Botha, {\it Exposure render: An interactive
  photo-realistic volume rendering framework},  {\em PLoS ONE} {\bf 7} (2012)
  e38586.

\bibitem{CTartifact}
J.~F. Barrett and N.~Keat, {\it {{A}rtifacts in {C}{T}: recognition and
  avoidance}},  {\em Radiographics} {\bf 24} (2004), no.~6 1679--1691.

\end{thebibliography}\endgroup

\end{document}